\documentclass[a4paper,11pt]{article}
\usepackage{pos}
\usepackage{bm}
\usepackage{xspace}
\usepackage{siunitx}

\def\ga{g_{a\gamma}}

\def\lat{\textit{Fermi/LAT}\xspace}

\title{AGN spectral variability across activity states and searches for axion-like particles}
\ShortTitle{AGN variability and ALP searches}

\author*[a]{Denys Malyshev}
\author[b,c]{Lidiia Zadorozhna}
 \author[d]{Yuriy Bidasyuk}
\author[a]{Andrea Santangelo}
\author[b]{Oleg Ruchayskiy}

\affiliation[a]{Institut f{\"u}r Astronomie und Astrophysik T{\"u}bingen, Universit{\"a}t T{\"u}bingen,\\
  Sand 1, D-72076, T{\"u}bingen, Germany}

\affiliation[b]{Niels Bohr Institute, University of Copenhagen,\\
Jagtvej 155A, 2200, Copenhagen, Denmark}

\affiliation[c]{Department of Quantum Field Theory and Astroparticle Physics, Faculty of Physics, Taras Shevchenko National University of Kyiv,\\
Hlushkova ave. 4, 03127, Kyiv, Ukraine}
\affiliation[d]{Bogolyubov Institute for Theoretical Physics
 of the National Academy of Sciences of Ukraine, Metrolohichna str. 14-b, 03143, Kyiv, Ukraine}

\emailAdd{denys.malyshev@astro.uni-tuebingen.de}
\emailAdd{zadorozhna@nbi.ku.dk}
\emailAdd{bidasyuk@bitp.kyiv.ua}
\emailAdd{andrea.santangelo@uni-tuebingen.de}
\emailAdd{oleg.ruchayskiy@nbi.ku.dk}

\abstract{Axion-like particles (ALPs) are compelling candidates for dark matter and potential portals to new physics beyond the Standard Model. Photons traversing magnetized regions can convert into ALPs, producing characteristic, energy-dependent absorption features in astrophysical spectra. The probability of such conversions depends sensitively on both the photon energy and the properties of the intervening magnetic fields.

Most existing searches have focused on individual astrophysical sources, but uncertainties in the structure and strength of cosmic magnetic fields have limited their reach. Recently, we have demonstrated that active galactic nuclei (AGNs) observed through galaxy clusters provide especially promising targets for ALP searches. By stacking multiple AGN-cluster sightlines, one can average over poorly known magnetic field configurations in galaxy clusters and recover a distinctive ALP-induced spectral suppression, thereby significantly enhancing sensitivity.

In this work, we investigate a possible systematic uncertainty in such analyses: the intrinsic time-variability of AGN spectra. We demonstrate that AGN flux variability is correlated with spectral hardness, and that time-averaging over flaring and quiescent states can potentially mimic the suppression features imprinted by ALP-photon mixing. Our findings imply that the recent constraints remain conservative, and that incorporating detailed spectral variability into stacking analyses can further sharpen the search for axion-like particles.}

\FullConference{Multifrequency Behaviour of High Energy Cosmic Sources - XV (MULTIF2025)\\
 9-14 June, 2025\\
Mondello, Palermo, Italy\\}

\begin{document}
\maketitle
%\linenumbers

\section{Introduction}
Axion-like particles (ALPs) are hypothetical pseudo-scalar particles predicted in many extensions of the Standard Model \cite{Jaeckel:2010ni,Choi:2020rgn}. They can exist across a wide range of masses while coupling only extremely weakly to ordinary matter~\cite{DiLuzio:2020wdo}. Because of these properties, ALPs are considered both leading dark matter candidates and potential mediators connecting the observed particles to a broader \emph{dark sector}~\cite{1983PhLB..120..127P, 1983PhLB..120..133A, 1983PhLB..120..137D, Alekhin:2015byh,Chadha-Day:2021szb}. An extensive experimental effort has been pursued for decades~\cite{Irastorza:2018dyq}, yet significant regions of well-motivated parameter space remain unexplored~\cite{AxionLimits}.

A key feature of ALPs is their mixing to photons in external magnetic fields:
\begin{equation}
    \label{eq:lagrangian}
    \begin{aligned}
  {\cal L}=-\frac{1}{4}
  F_{\mu\nu}{F}^{\mu\nu}+\frac{1}{2}\left(\partial_{\mu}a\partial^{\mu}a-m_{a}^2a^2\right)+\frac{1}{4}\ga F_{\mu\nu}\tilde{F}^{\mu\nu}\,a.
\end{aligned}
\end{equation}
Here, $a$ is the ALP field, $m_a$ is its mass, $F_{\mu\nu}$ is the electromagnetic field strength tensor, and $\tilde{F}_{\mu\nu}\equiv\frac{1}{2}\varepsilon_{\mu\nu\rho\sigma} F^{\rho\sigma}$ is the electromagnetic dual tensor. The photon-ALP coupling constant, $\ga$, characterizes the interaction strength.

This coupling allows photons to oscillate into ALPs with probability growing with both magnetic field strength and the distance traveled. Astrophysical environments provide extremely long baselines and coherent magnetic domains, enabling astrophysical searches to probe couplings far below the reach of current laboratory experiments~\cite{AxionLimits} .
Photon-ALP mixing can generate distinctive, energy-dependent spectral modulations superimposed on otherwise smooth $\gamma$-ray spectra of background photon sources. The unambiguous detection of these features would constitute a clear signature of ALPs and thus strongly motivate continued observational efforts.
The morphology of ALP-induced features is highly sensitive to uncertain magnetic field configurations, e.g. orientation of the magnetic field and/or spatial scales on which the magnetic field remains approximately constant.
Thus, these studies require marginalisation over the possible field configurations.
This necessity reduces sensitivity, leaving large parts of parameter space unconstrained.

\begin{figure}[!t]
    \centering
    \includegraphics[width=0.49\linewidth]{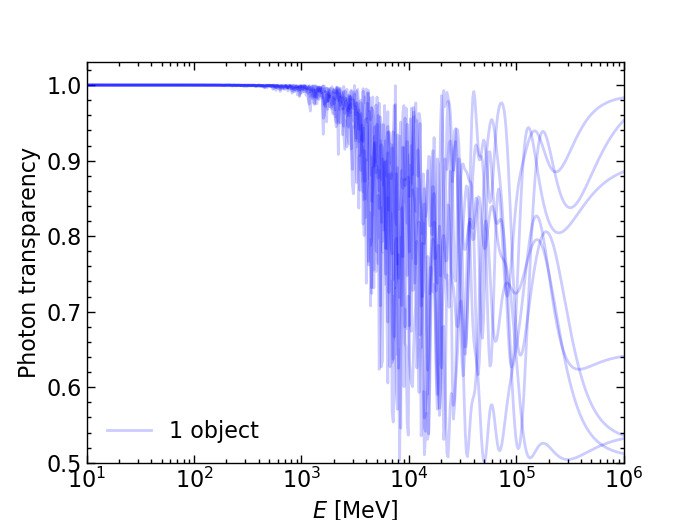}~
    \includegraphics[width=0.49\linewidth]{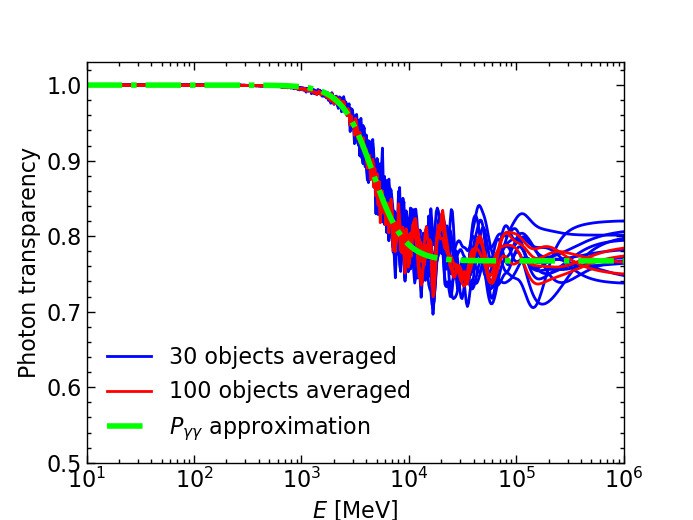}
    \caption{\textbf{Photon survival probability and its averages.} \textit{Left:} Photon transparency (survival probability $P_{\gamma\gamma}$) for photons traversing a galaxy cluster, shown for different realisations of the cluster magnetic field with identical global characteristics (blue lines). The turbulent magnetic field is modelled as a sequence of cells with constant but randomly oriented fields along the photon trajectory. The transparency has a complex shape, with little similarity between different realizations.\\
      \textit{Right panel:}
      {\bf Black lines} demonstrate the effect of averaging over 30 (blue) or 100 (red) randomly selected realizations.
      The self-similar shapes emerge, with the
      \textbf{green dot-dashed line} showing the asymptotic analytical approximation to these lines (Equation~\protect\ref{eq:p_ga}).
    The ALP parameters for all curves are the same.}
    \label{fig:conversion}
\end{figure}

Our recent  paper~\cite{Malyshev:2025iis} introduced a strategy that tackles this challenge.
Rather than analyzing individual objects, we exploited large statistical samples and stacked their spectra. Such a method enhanced the \emph{predictability of the signal}, converting quasi-random absorption patterns into a \textbf{distinct, step-like spectral suppression} largely independent of magnetic field uncertainties (Figure~\ref{fig:conversion}). In the limit of many objects, the energy-dependent photon transparency can be approximated by a simple smooth analytic function% takes well-defined shape
\begin{equation}
P_{\gamma\gamma}(E) %\equiv 1 - \bigl\langle P_{\gamma a}(E)\bigr\rangle
= 1 - \frac{p_0}{1 + (E_c/E)^k}.
\label{eq:p_ga}
\end{equation}
Here, $p_0$ specifies the suppression depth at high energies, $E_c$ the characteristic transition energy, and $k$ the sharpness of the transition.

Using the latest \lat\ AGN catalog~\cite{2022ApJS..260...53A,4fgldr4},  we identified 32 sources located behind known galaxy clusters \cite{2018MNRAS.475..343W,planck_2015sz,erass1_clusters}.
The analysis of~\cite{Malyshev:2025iis} used time-averaged AGN spectra reported in the catalog.
Such spectra were well-described by a EBL-absorbed log-parabola model -- a smooth three-parameter function, against which we sought the suppression~\eqref{eq:p_ga}.
This procedure allowed us to obtain strongest to-date bounds for ALP in the nano-eV mass range.

However, the AGNs are known to be strongly time-variable objects with the spectral parameters changing with the flux of the source.
Indeed, during flares, the spectra can appear harder \cite{2010ApJ...714L..73A, 2021MNRAS.504.1103R}. This may reflect the change of the intrinsic emission mechanism of the AGN (or may be an artifact of the limited photon statistics). 
The averaging over different spectral states can therefore produce a spectrum that cannot be described by a simple model; for example, averaging power-law spectra with different indices generally results in a spectrum that departs from a pure power law.
Consequently, spectral residuals arising from an improperly chosen baseline model can be mistaken for a photon-ALP conversion feature. In this work, we investigate this effect in detail and assess the extent to which AGN spectral variability can influence the derived ALP bounds.

\begin{figure}[!t]
    \centering
    \includegraphics[width=0.75\textwidth]{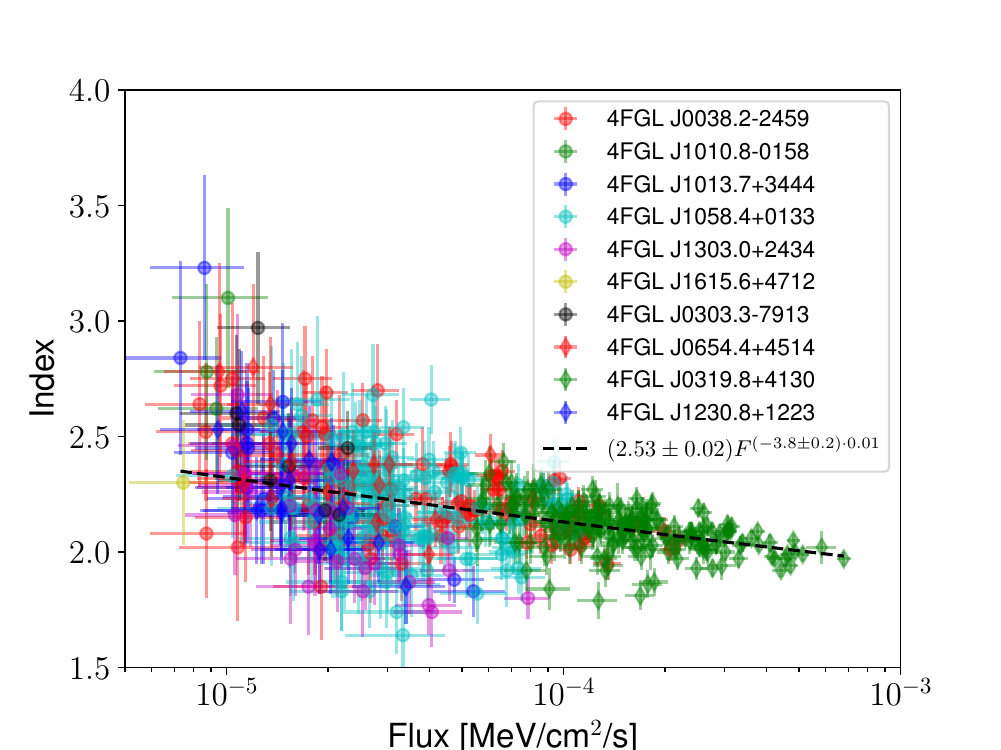}
    \caption{\textbf{Correlation between photon index $\Gamma$ and flux $F$ (0.1-100~GeV) in monthly bins for ten AGNs.} Points of different colours correspond to different sources as indicated in the legend; statistical uncertainties are shown as error bars in both coordinates. The black dashed line represents the best-fit power-law model to the combined data, with the fitted parameters given in the legend.}
    \label{fig:index_flux_correlation}
\end{figure}

\section{Variability effects}

In order to investigate the correlation between the AGN state and spectral shape, we took 10 AGNs from our study \cite{Malyshev:2025iis} with light curves present in \lat Light Curve Repository\footnote{See \href{https://fermi.gsfc.nasa.gov/ssc/data/access/lat/LightCurveRepository/index.html}{\lat Light Curve Repository webpage}. Please note that only variable sources (variability index $>
21.67$ in 4FGL-DR2 catalogue) are reported in LCR.} (LCR) \cite{2023ApJS..265...31A}. In addition to light curves, covering the period since 2008 and continuously updated, this repository provides best-fit spectral parameters derived from the standard likelihood analysis in the 0.1–100~GeV energy band, as well as fluxes for each time bin (binned in days, months, etc.).
For each source, we retrieved monthly- and weekly-binned data, including photon flux $F$, photon index $\Gamma$, and their statistical uncertainties.  
The data were filtered  considering only periods of significant detection ($\Delta\Gamma<\Gamma/3$,  $\Delta F<F/3$, $\Gamma<3.5$), and additionally require fit convergence (via  $F<10^{-2}$~MeV\,cm$^{-2}$\,s$^{-1}$ and $\Delta\Gamma\neq 0$  ), where $\Delta F$ and $\Delta\Gamma$ are the statistical uncertainties on the flux $F$ and photon index $\Gamma$.

To search for potential systematics, we examined the flux -- photon-index relation (Fig.~\ref{fig:index_flux_correlation}). We find a clear anti-correlation: brighter, flaring states exhibit harder spectra (smaller photon index), whereas quiescent periods are softer (larger photon index). This trend may introduce systematics that can mimic a photon--ALP conversion-like signal.

To further quantify the spectral slope-flux ($\Gamma-F$) correlation we consider a simple power-law dependency of these quantities:

\begin{equation}
\label{eq: g(f)}
    \Gamma(F) = A \left( \frac{F}{10^{-6}\mbox{\,MeV\,cm}^{-2}\mbox{\,s}^{-1}} \right)^B ,
\end{equation}
where $F$ is the photon flux in the 0.1--100~GeV range in MeV\,cm$^{-2}$\,s$^{-1}$, and $A$ and $B$ are free fit parameters. We derived these parameters via fitting the monthly-binned light curve points, see black dashed line in Fig.~\ref{fig:index_flux_correlation}. The asymptotic $1\sigma$ statistical uncertainties on the derived parameters are presented in the legend of Fig.~\ref{fig:index_flux_correlation}.

\begin{figure}[t]
    \centering
    \begin{minipage}{0.49\textwidth}
        \centering
        \includegraphics[width=\linewidth]{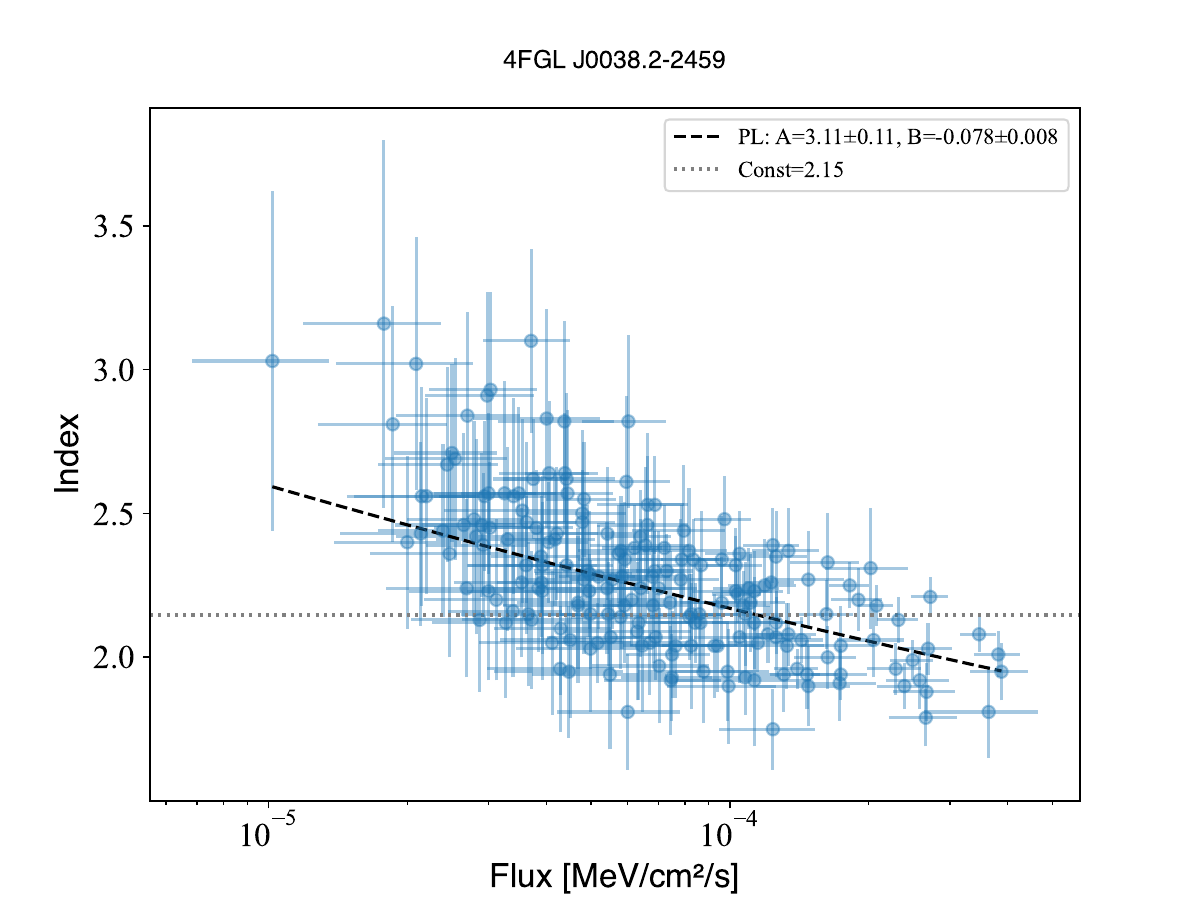}
    \end{minipage}
    \hfill
    \begin{minipage}{0.49\textwidth}
        \centering
        \includegraphics[width=\linewidth]{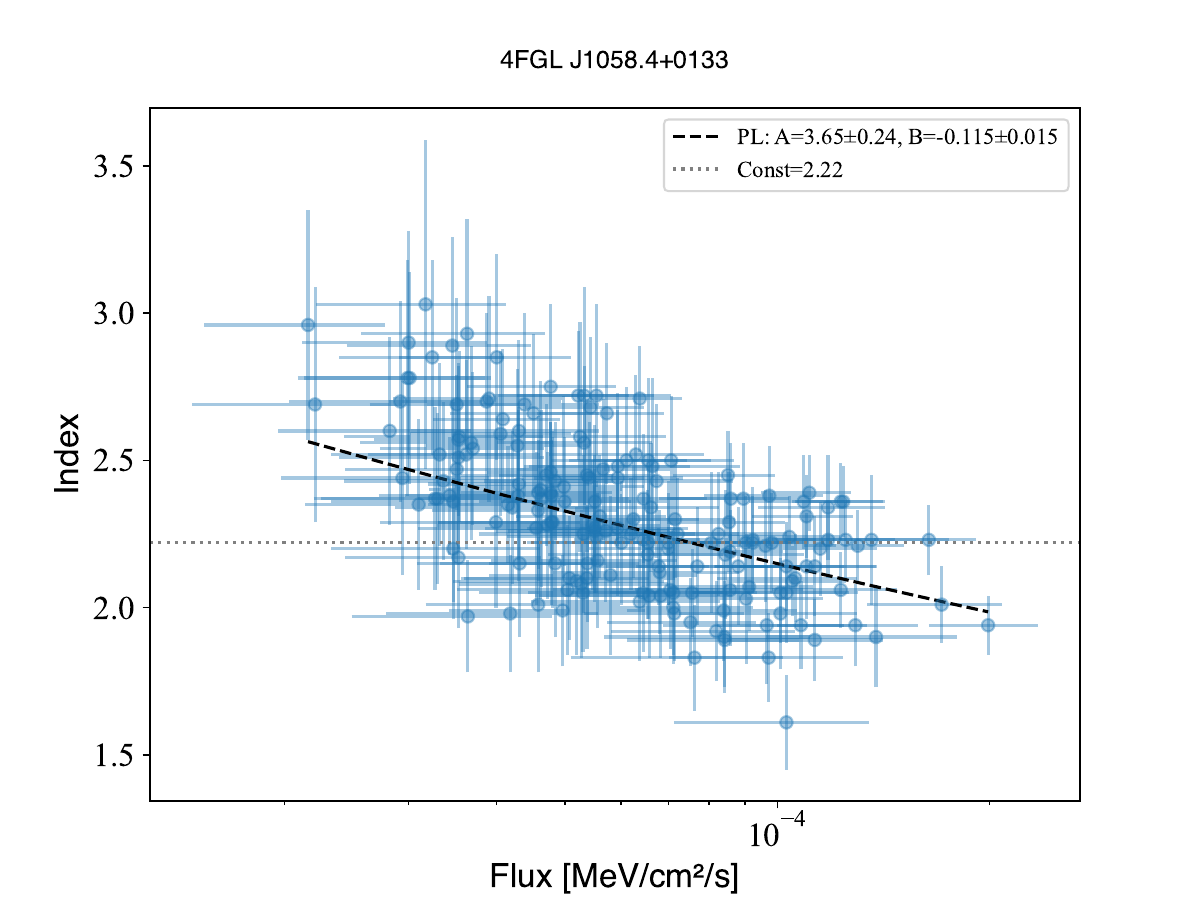}
    \end{minipage}
    \caption{
        \textbf{Weekly-binned relation between the photon index $\Gamma$ and photon flux $F$ for two AGN from our sample, selected due to their extensive observational statistics.} Blue points represent the Fermi-LAT measurements with statistical uncertainties (error bars) on both flux and photon index. Black dashed lines show the best-fit power-law models, while gray dotted lines indicate the best-fit constant indices. The corresponding best-fit parameters (see Eq.~\eqref{fig:gamma_flux}) and statistical uncertainties are listed in each panel.}
    \label{fig:gamma_flux}
\end{figure}

The observed correlation appears not only in the combined all-AGN dataset but also in individual sources with sufficiently high photon statistics (see Fig.~\ref{fig:gamma_flux}) and on weekly time scales. For illustration, we show $\Gamma(F)$ for two bright AGNs: 4FGL~J0038.2$-$2459 (QSO B0035-252) and 4FGL~J1058.4+0133 (QSO B1055+018). To assess the significance, we compare the best-fit $\chi^2$ values for a constant-slope model, $\Gamma(F)=\text{const}$, and a power-law dependence, Eq.~\ref{eq: g(f)}. The difference, $\Delta\chi^2=\chi^2_{\rm Const}-\chi^2_{\rm PL}$, indicates that the power-law model is favored, with a statistical significance of $N_\sigma=10.22\sigma$ for 4FGL~J0038.2$-$2459 and $N_\sigma=7.46\sigma$ for 4FGL~J1058.4+0133.

\begin{figure}
    \centering
    \includegraphics[width=0.75\linewidth]{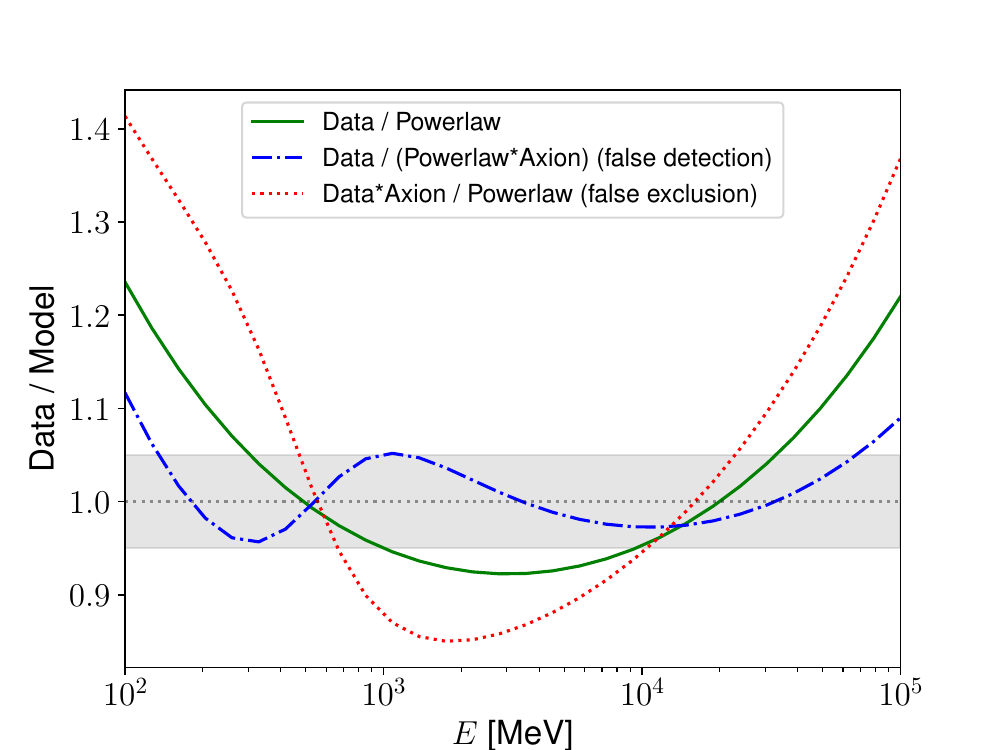}
    \caption{
    %\textbf{Potential false-detection of ALPs due to flux-index correlation.} The objects-averaged ratio of the simulated data spectra to the baseline power-law model (solid green line) and to the photon-ALP conversion modified model (blue dot-dashed line, ``false ALP detection'' case). The red dotted line (``false exclusion'' case) correspond to the ratio of the photon-ALP modified data spectra to the baseline poweraw model. The typical systematic uncertainty of \lat is shown with a gray shaded region. The lines providing the better fit are located closer to 1, see text for the details.
    {\textbf{Potential false detection and false exclusion of ALPs due to flux-index correlation.} This figure demonstrates that the index-flux correlation can induce spurious detection but cannot hide a genuine ALP signal, if one were present. The solid green line represents the ratio of simulated data to the best-fit power-law model. If the power-law index were independent of the flux, this ratio would be equal to unity within statistical uncertainties. However, the figure clearly demonstrates a significant deviation of the green curve from a horizontal line, well outside the nominal $\pm 5\%$ uncertainty region. If we modify the best-fit model to include ALP-induced suppression, the ratio becomes closer to unity (blue dot-dashed line), except in the lowest and highest energy bins, where systematic uncertainties are larger. This indicates that ALP-like spectral features can spuriously emerge from the correlation bias alone, as including them in the model artificially "improves" the quality of fit. Conversely, the red dotted line ("false exclusion" case) displays the ratio of data containing a genuine ALP signal to a pure power-law model. 
    This ratio deviates more strongly from unity than the green curve, indicating that the ALP signal remains detectable despite the bias. 
    }}
    \label{fig:false_detection}
\end{figure}

\section{Discussion: variability effects on ALP bounds}
\label{sec:ALPs_bounds}

The performed analysis indicates the presence of a negative correlation between photon index and flux -- that is, whether higher flux states systematically correspond to harder spectra. Although this correlation was observed earlier in the GeV band for several bright AGNs~\cite{2010ApJ...714L..73A, 2021MNRAS.504.1103R} we cannot exclude also the possible data-analysis bias connected to limited photon statistics. For a relatively small flux the number of photons detected from a source at high energies is low, forcing the fitting procedure to converge to softer indexes as for these indexes one can expect generally low number of high-energy photons.
{We note in passing that Cherenkov telescopes (such as HESS or the forthcoming CTAO) can only partially alleviate this situation, since ground-based observations are constrained to relatively short exposure times for each individual AGN. In addition, the limited sensitivity of ground-based instruments requires longer exposure times to achieve a detection and therefore reduces the effectiveness of simultaneous GeV/TeV observations and the associated data analysis.
A wide-field instrument such as HAWC~\citep{2023NIMPA105268253A} (or the future SWGO~\citep{2025arXiv250601786S}), which monitors a substantial fraction of the celestial hemisphere, could in principle provide more continuous coverage; however, the number of AGNs observed by HAWC remains very limited~\citep{2021ApJ...907...67A,2024icrc.confE.805U}. Finally, in the TeV energy range, absorption by the extragalactic background light becomes significant, introducing additional systematic uncertainty into precise measurements of the spectral index~\citep{2013APh....43..112D}.}

While detailed studies of the origin (physical or apparent bias) of $\Gamma-F$ correlation is beyond the scope of this work, we note that this correlation can mimic the signal expected from photon-ALP conversion -- suppression of the high-energy part of spectrum. 
To illustrate this we consider the best-fit spectral power-law parameters of the objects reported in LCR for monthly time bins. Based on these parameters we built individual power-law model-spectra for each time bin and calculated then the time-averaged spectrum for each object $S_{\rm obj}(E) = \langle S_{\rm obj}(E,t)\rangle_t$.

Following~\cite{Malyshev:2025iis}, we analyzed the obtained spectra by comparing a baseline power-law model with the same model modified by possible photon-ALP conversion (see Eq.~\eqref{eq:p_ga}). We first fit the time-averaged spectrum of each object "$\rm obj$" with a baseline power-law model $M_{\rm obj}(E)$ and then consider the quantity $\langle S_{\rm obj}(E)/M_{\rm obj}(E)\rangle_{\rm obj}$, averaged over all AGNs. This represents the average ratio, across objects, of the simulated time-averaged data points to the baseline model, and is shown as the green solid line in Fig.~\ref{fig:false_detection}. 
We note that this line exhibits a systematic trend, generally lying outside the typical $\pm 5\%$ systematic uncertainty of \lat, indicated by the gray shaded region.

We next considered an analogous quantity for a photon–ALP absorption–corrected model, $M'_{\rm obj}(E) = M_{\rm obj}(E) P_{\gamma\gamma}(E)$, where $P_{\gamma\gamma}(E)$ is given by Eq.~\eqref{eq:p_ga} for an ALP mass within the marginal $2\sigma$ detection region of~\cite{Malyshev:2025iis}, i.e.\ $m_a = 10^{-9}$ eV and\footnote{For illustration, we adopt $g_{a\gamma}$ larger by a factor of 1.5 than the best-fit $g_{a\gamma}$ from~\cite{Malyshev:2025iis}.} $g_{a\gamma} = 3 \cdot 10^{-12}$ GeV$^{-1}$.
The corresponding averaged ratio, $\langle S_{\rm obj}(E)/M'_{\rm obj}(E)\rangle_{\rm obj}$, is shown in Fig.~\ref{fig:false_detection} as a blue dot-dashed line. This curve stays much closer to unity than in the baseline model and generally lies within the typical \lat systematic uncertainty across most energies. This behavior indicates an overall better fit of the data with the photon–ALP absorption model compared to the baseline power law. In the analysis of real data, such an improvement would correspond to a case of “false ALP detection.”

Contrary, our attempt to inject photon-ALP conversion to the data spectra and fit the obtained spectra with the baseline power-law model resulted in a similar to discussed-above ratio $\left\langle\langle S_{\rm obj}(E,t)P_{\gamma\gamma}(E)\rangle_t / M_{\rm obj}(E)\right\rangle_{\rm obj}$ shown with red dotted line in Fig.~\ref{fig:false_detection}. This line is further from 1 even in comparison to the greeng one which corresponds to the worse fit of the photon-ALP conversion-affected data with the baseline power-law model. We conclude that the flux-slope dependency can potentially lead to the false-detection of ALPs but unlikely to cause the false-exclusion problem.

While the flux-slope correlation can mimic an ALP signal, this effect can be mitigated. In our analysis of the AGN spectra dataset \cite{Malyshev:2025iis}, we applied different baseline models to individual objects, but used averages over 14 years of data.
Because the spectral shape depends on the object’s flux, we now propose splitting each object’s dataset into intervals of approximately constant flux. For each such interval, a spectrum should be extracted and fitted with its own baseline model. In other words, the baseline modeling must be performed independently for every object and every flux level.
Such a flux-based binning enables us to distinguish states with different intrinsic spectral shapes and to assign an appropriate model to each state.  Grouping data by flux provides a clearer separation between intrinsic variability and possible new physics signatures.

\section*{Acknowledgments}

The authors acknowledge support by the state of Baden-W\"urttemberg through bwHPC.  
This project has received funding through the MSCA4Ukraine project, grant number Ref 1.4 - UKR - 1245772 - MSCA4Ukraine, which is funded by the European Union. Views and opinions expressed are however those of the author(s) only and do not necessarily reflect those of the European Union, the European Research Executive Agency or the MSCA4Ukraine Consortium. Neither the European Union nor the European Research Executive Agency, nor the MSCA4Ukraine Consortium as a whole nor any individual member institutions of the MSCA4Ukraine Consortium can be held responsible for them. This work was partially supported by the Ministry of Education and Science of Ukraine under the project “Search for dark matter and particles beyond the Standard Model” (0125U002260).
This research was conducted with support from the Centre for the Collective Use of Scientific Equipment "Laboratory of High Energy Physics and Astrophysics" of Taras Shevchenko National University of Kyiv.

\bibliographystyle{JHEP}
\bibliography{proceedings}
% \begin{thebibliography}{99}
% \bibitem{...}
% ....

% \end{thebibliography}

\bigskip
\bigskip
\noindent {\bf DISCUSSION}
\bigskip\newline
\noindent {\bf S\"olen Balman:} Can you put some constraint for dark matter densities using your method?
\bigskip\newline
\noindent {\bf Denys Malyshev:} Not really, our method is not relying on the dark matter densities in clusters of galaxies, so can not be directly used to probe this quantity. Still our limits (assuming zero systematics of \lat) are relevant for the ALP parameters region where ALPs can present up to all dark matter in the Universe. In this sense our limits are relevant for ALP dark matter.

\end{document}